\documentclass[%
prl%
 ,twocolumn
 ,secnumarabic%
,amssymb, amsmath,nobibnotes, aps, pre]{revtex4-2}
\usepackage[dvipdfm]{graphicx}
\usepackage{latexsym,mathrsfs}
\usepackage{bm}% bold math

\def \av#1{\langle #1\rangle}

\begin{document}

\title{Dynamical toy model of interacting $N$ agents robustly exhibiting Zipf's law}

\author{Tohru Tashiro}
\affiliation{Comprehensive Education Center, Aichi University of Technology, 50-2 Manori, Nishihasama-cho, Gamagori, Aichi 443-0047, Japan}
\author{Megumi Koshiishi,  Tetsuo Deguchi}
\affiliation{Department of Physics, Ochanomizu University, 2-1-1 Ohtsuka, Bunkyo, Tokyo 112-8610, Japan}

\date{\today}

\begin{abstract}
We propose a dynamical toy model of agents which possess a quantity and have an interaction radius depending on the amount of the quantity.
They exchange the quantity with agents existing within their interaction radii.
It is shown in the paper that the distribution of the quantity of agents is robustly governed by Zipf's law for a small density of agents independent of the number of agents and the type of interaction, despite the simplicity of the rules.
The model can exhibit other power laws with different exponents and the Gaussian distributions.
The difference in the mechanism underlying Zipf's law and other power laws are studied by mapping the systems into graphs and investigating quantities characterizing the mapped graph.
Thus, this model suggests one of the origins of Zipf's law, i.e., the most common fundamental characteristics necessary for Zipf's law to appear.  
\end{abstract}

\maketitle

%\section{Introduction}

Various phenomena characterized by power law distributions ubiquitously occur in nature, e.g.,  a size of microorganisms \cite{Jonasz11,Gaedke92,Quinones03,Baho19}, a stress drop in the stick-slip instability \cite{Lebyodkin00,Ananthakrishna01,Chihab03,Ananthakrishna04,Lebedkina08,Lebyodkin12,Gibiansky13}, a link in the Internet \cite{Adamic00}, and the number of chromosome aberrations in cancer \cite{Frigyesi03}.

The following statistical law related to power law is well-known:
for ranked $N$ observations about a quantity, the expression
\begin{equation}
m(f)=m_{\rm max}f^{-\beta}
\label{eq:1}
\end{equation}
holds, where  $m(f)$ is the observed value of the quantity at the $f$th rank $(1\le f\le N)$, $m_{\rm max}$ means the largest value of the observation, and $\beta\simeq1$.
The empirical law is referred to as {\it Zipf's law}.

Solving Eq.~(\ref{eq:1}) for $f$ and dividing it by $N$ yield
$
f(m)/N = {m_{\rm max}}^{1/\beta}m^{-1/\beta}/N
$.
For sufficient large $N$, we can interpret the relation as the ratio of the observation of the value greater than $m$, i.e., the cumulative distribution function (CDF):
$P(>m) = \int_{m}^{m_{\rm max}}p(m'){\rm d}m'
 ={m_{\rm max}}^{1/\beta}m^{-1/\beta}/N
$, where $p(m)$ is the probability density function (PDF). 
Thus, by differentiating the equation by $m$, we can obtain
$
p(m) = {m_{\rm max}}^{1/\beta}m^{-1/\beta-1}/{\beta N}
$.
That is, Zipf's law can be regarded as a power law distribution whose exponents of CDF and PDF are nearly equal to $-1$ and $-2$, respectively.
We can also find the law or the power law distribution in many different fields, e.g., words in novels \cite{Montemurro02}, populations in cities \cite{Gabaix99,Gabaix99-2,Kuninaka08,Ghosh14}, income of companies \cite{Okuyama99,Axtell01}, and magnitudes of earthquakes \cite{Rundle18}.
The fact that the power law distribution with the universal exponent can be observed cross-sectionally indicates the existence of a common mechanism behind the universality.

After Simon's pioneering work \cite{Simon55}, various studies have been proposed to derive the universal law, e.g., agent-based models \cite{Mansury07,Cirillo12} and random multiplicative processes \cite{Levy96,Takayasu97}.

In an agent-based model with discretized times proposed by Gaffeo et. al. \cite{Gaffeo08}, there exist synthetic firms, workers/consumers, and banks. The firms determine the amount of product to be produced and the price. The workers can apply to one of the firms which supplies the highest wage as far as they know at each period. If an internal financial resource of a firm is not enough for paying wages, the firm can borrow funds from one of the banks. Each consumer buys the product from one of the firms whose price is not necessarily the lowest. 

It is clear how the agents interact with each other in the model, but the model is so specific to economic activity that it is insufficient to elucidate the origins of the universality of Zipf's law.
By contrast, it is quite difficult to understand what kind of interactions among elements of a system exhibiting the law exists behind the multiplicative noise of the random multiplicative processes. 

In the paper, we propose a dynamical toy model of agents which interact each other by simple rules abstracted from phenomena that exhibit Zipf's law.
The agents possess an amount of {\it quantity} (e.g., which respectively represents money and people if the agent is a firm and a city) and exchange a portion of quantity according to the rules at each discretized time step.
The distribution of quantity at a small density of agents is equivalent to Zipf's law.
It is found that the dynamical toy model shows the power law distribution with other exponents by changing the density of agents.
Moreover, the Gaussian distribution also can be derived when the agents interact with each other at short range. 
Thus, this study unveils one of the origins of the universality of Zipf's law.

%\section{Toy model with interacting $N$ agents}
%\label{toymodel}

What are the common fundamental characteristics for the interaction among elements of the systems that exhibit  Zipf's law?
Indeed, we might find several features. However, let us here employ the following one: {\it the more amounts of quantity the elements have, the more elements they are able to interact with.} 
This seems intuitively plausible, and we can find evidences to support it.
It has been reported in Ref.~\cite{Watanabe13} that the sales and the number of business partners also have positive correlation for approximately 500000 Japanese companies.
In addition, it has been theoretically predicted that the number of earthquakes caused by a mainshock increases exponentially with the magnitude of it \cite{Utsu71,Ogata88,Passarelli18}.

We now construct a dynamical toy model of $N$ agents interacting with each other reflecting their properties.
The agents exist on the two-dimensional $L\times L$ lattices with length one, 
and the boundary conditions are periodic.
They have an amount of quantity $m$ and an interaction radius $r(m)$ which is a monotonically increasing function of $m$. 
Let us denote the amount of quantity of the $i$th agent by $m_i$.
In addition, we shall define the density of agents by $d\equiv N/L^2$.
The quantity of the agents changes with time step by the following rules:
(i) Set all the amounts of quantity one at the initial time step, $t=0$.
(ii) If the $i$th agent stays within the interaction radius of the $j$th agent, exchange $\Delta m = \gamma\times {\rm min}(m_i, m_j)$, where we call $\gamma$ an {exchange coefficient}.
The $i$th or $j$th agent obtains $\Delta m$ with probability $1/2$.
(iii) If an amount of quantity of an agent becomes less than one after all the interactions at the time step are performed, set the value one at the next time step.
(iv) Move all the agents to one of the nearest lattices randomly and return to (ii) with increasing the time step $t$ by one.

%\section{numerical results}

\begin{figure}[h]
  \begin{center}
\includegraphics[scale=0.75]{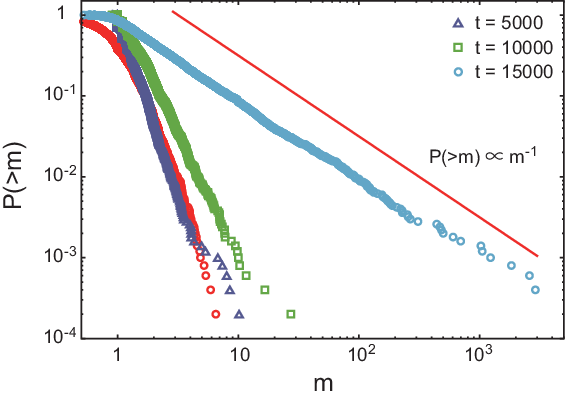}\vspace{1pc}
    \caption{\label{fig:exCumPDF}The cumulative distribution function, $P(>m)$, with $N=5000$, $\gamma=0.1$, $d=2^{-4}~\%$, and $r(m)=m$ is plotted in a log-log scale at $t=5000$ (navy triangles), 10000 (green squares), and 15000 (blue circles). 
For comparison, $P(>m)$ for the same parameters, without the rule (iii), at $t=15000$ is plotted by red circles. The red line represents $P(>m)\propto m^{-1}$.
}
  \end{center}
\end{figure}

We have run numerical simulations of the model with varying $d$, $\gamma$, $N$, and $r(m)$, and then calculated CDF for observing quantity that agents possess.
Figure~\ref{fig:exCumPDF} shows CDF with $N=5000$, $\gamma=0.1$, $d=2^{-4}~\%$, and $r(m)=m$ in a log-log scale at $t=5000$ (navy triangles), 10000 (green squares), and 15000 (blue circles). $P(>m)\propto m^{-1}$ is also drawn as a supporting line in the figure, so that we can see that CDF is a power law function and the exponent is getting close to $-1$ as $t$ increases.
After $t$ is larger than approximately 15000, CDF changes very little, which implies that there exists a stationary state where the exponent of CDF is almost $-1$: Zipf's law is satisfied.
For comparison, CDF for the same parameters at $t=15000$, without the rule (iii), is plotted by red circles.
The result indicates that the rule is necessary for leading to the power law.

\begin{figure*}[ht]
    \includegraphics[scale=0.63]{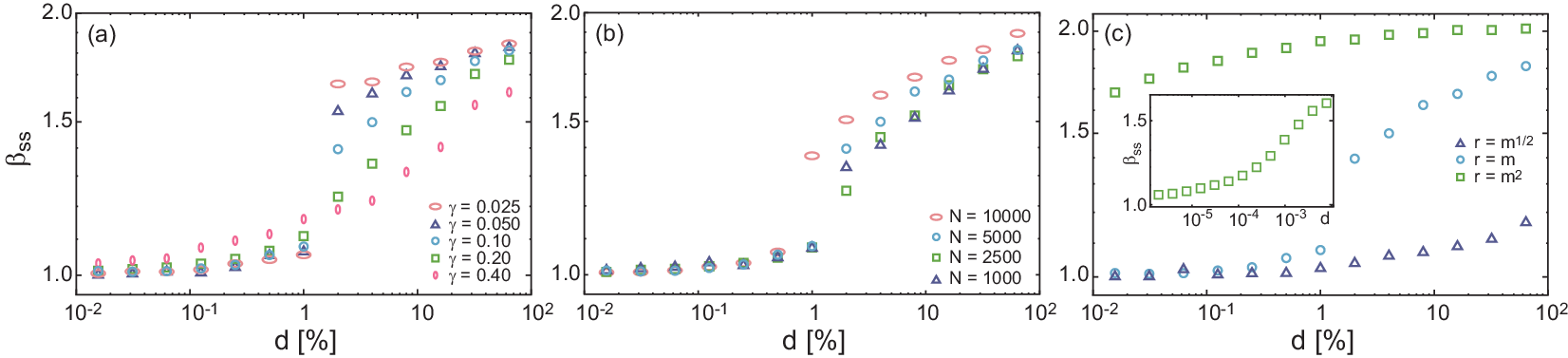}\hspace{2pc} 
    \caption{\label{fig:merge_4_5_6}The dependence of $\beta_{\rm SS}$ on $d$ is plotted by a log-log scale for (a) $\gamma = 0.025$ (pink horizontal ellipses), $0.050$ (navy triangles), $0.10$ (blue circles), $0.20$ (green squares), and $0.40$ (pink vertical ellipses) with $N=5000$ and $r(m) = m$,
(b) $N = 1000$ (navy triangles), $2500$ (green squares), $5000$ (blue circles), and $10000$ (pink horizontal ellipses) with $\gamma=0.10$  and $r(m)=m$, and 
(c) $r(m)=m^{1/2}$ (navy triangles), $m^{2}$ (green squares), and $m$ (blue circles) with $\gamma=0.10$ and $N=5000$.
}
\end{figure*}

Let us fit the datasets of CDF by $P(>m) = {\rm const.}\times m^{-1/\beta}$ where $\beta$ is the same one of Eq.~(\ref{eq:1}).
We denote the best fit $\beta$ at the stationary state as $\beta_{\rm SS}$.
 Figure~\ref{fig:merge_4_5_6}(a) shows $\beta_{\rm SS}$ for $\gamma = 0.025$ (pink horizontal ellipses), $0.050$ (navy triangles), $0.10$ (blue circles), $0.20$ (green squares), and $0.40$ (pink vertical ellipses) with $N=5000$ and $r(m)=m$ as a function of $d$ by a log-log scale.
It is clarified from the figure that $\beta_{\rm SS}$ is getting closer to one, which means that Zipf's law holds, regardless of the value of the exchange coefficient with decreasing $d$ and it drastically changes centered at $d=1~\%$ for the small value of the exchange coefficient.

We now explore how the number of agents affects $\beta_{\rm SS}$.
Figure~\ref{fig:merge_4_5_6}(b) shows $\beta_{\rm SS}$ as a function of $d$  in a log-log scale for $N = 1000$ (navy triangles), $2500$ (green squares), $5000$ (blue circles), and $10000$ (pink horizontal ellipses) with $\gamma=0.10$ and $r(m) = m$.
Even though the transition density where $\beta_{\rm SS}$ changes to one for $N=10000$ differs from that for the other values of $N$ a little,
$\beta_{\rm SS}$ at small $d$ behaves similarly regardless of $N$: again, Zipf's law holds for the small density.

Let us investigate the behaviors of $\beta_{\rm SS}$ for other types of interaction radius.
In Fig.~\ref{fig:merge_4_5_6}(c), we plot $\beta_{\rm SS}$ as a function of $d$ in a log-log scale for $r(m)=m^{1/2}$ (navy triangles), $m^{2}$ (green squares), and $m$ (blue circles) with $\gamma=0.10$ and $N = 5000$.
We can see that $\beta_{\rm SS}$ for $r(m)=m^{1/2}$ approaches to one fastest, as $d$ decreases.
The values of $\beta_{\rm SS}$ for $r(m)=m^{2}$ is almost 1.68 for $d=2^{-6}~\%$. However, if $d$ becomes much smaller than the value, $\beta_{\rm SS}$ gets close to one, as shown in the inset of Fig.~\ref{fig:merge_4_5_6}(c).

%\section{Factors leading to the difference in the exponent}

Next, we study the difference in the mechanism underlying Zipf's law and other power law distributions.
We focus on the network structure of the model by mapping it into a random graph: we regard each agent as a {\it vertex} and assume that there is an {\it undirected edge} between two agents (vertices) if one belongs to the interaction radius of the other.
With $N = 5000$, $\gamma = 0.1$, and $r(m)=m$, the results for $d = 2^{-6}\%$ and $d = 2^6\%$ are employed as examples of Zipf's law and the other power law (the exponent of CDF is nearly equal to $-0.55$), respectively.

\begin{figure*}[t]
    \includegraphics[scale=0.59]{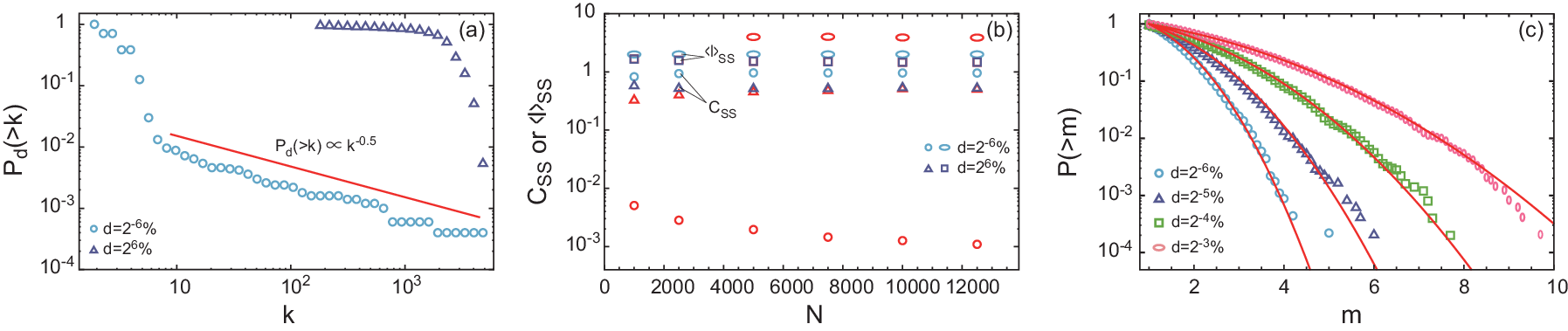}
     \caption{\label{fig:merge_7_8_10}(a) The cumulative degree distributions, $P_{\rm d}(>k)$, of mapped graphs for $d = 2^{-6}\%$ (blue circles) and $d = 2^6\%$  (navy triangles) with $N = 5000$, $\gamma = 0.1$, and $r(m)=m$ at $t=50000$. The red line represents $P_{\rm d}(>k) \propto k^{-0.5}$. (b) The mean clustering coefficients $C_{\rm SS}$ (circles and triangles) and the mean path length $\av{l}_{\rm SS}$ (horizontal ellipses and squares) of mapped graphs averaged over the stationary state as a function of the number of agents $N$ for $d = 2^{-6}\%$ (blue symbols) and $d = 2^6\%$  (navy symbols) with $\gamma = 0.1$ and $r(m)=m$.
For comparison, the clustering coefficients and mean path lengths of random graphs with the same number of vertices and the same average degrees of the mapped graphs with $d = 2^{-6}\%$ (red circles and horizontal ellipses, respectively) and $d = 2^{6}\%$ (red triangles and squares, respectively) are also plotted. (c) The cumulative distribution function, $P(>m)$, with $\Delta m=0.1$, $r(m)=0.5$, and $N=5000$ at $t=50000$ is plotted in a log scale for $d = 2^{-6}$ (blue circles), $2^{-5}~\%$ (navy triangles), $2^{-4}~\%$ (green squares), and $2^{-3}~\%$ (pink ellipses). The red curves represent $P_{\rm REFL}(>m)$.}
\end{figure*}

Figure~\ref{fig:merge_7_8_10}(a) shows the cumulative degree distributions, $P_{\rm d}(>k)$, of mapped graphs for $d = 2^{-6}\%$ (blue circles) and $d = 2^6\%$ (navy triangles) at $t=50000$ as a function of the degree, $k$.
From the figure, we can say that a large number of vertices in the mapped graph for $d = 2^{-6}\%$ have fewer than 10 connections
and that a part of the graph is scale-free, since the distribution exhibits power law behavior for $k > 10$ \cite{Barabasi16}.
The figure also shows that most of vertices in the graph for $d = 2^{6}\%$ have hundreds to thousands of degrees, indicating a more highly connected graph than that for $d = 2^{-6}\%$.
In fact, the connectances of the mapped graphs for $d = 2^{-6}\%$ and $d = 2^{6}\%$ at $t=50000$ are approximately equal to  $1.3\times10^{-3}$ and $4.7\times10^{-1}$, respectively. 

We can calculate a {\it critical interaction radius}, $r^{*}$, within which only one agent exists, for a given density $d$ as 
$r^{*} = 1/\sqrt{\pi d}$.
In the case that $r(m) = m$, the corresponding quantity, $m^{*}$, becomes $m^{*} = r^{*} = 1/\sqrt{\pi d}$.
Thus, an agent whose amount of quantity is less than $m^{*}$ has no interacting partners within the interaction radius.
If Zipf's law holds, the amount of quantity of an agent at $f$th rank can be expressed as $m(f) \simeq N/f$, because the amount of quantity at the lowest rank is one in the system.
Therefore, we can understand that the agents whose rank is greater than $f^* = N/m^* = N\sqrt{\pi d}$ have no partners in their radii.
For $N=5000$ and $d=2^{-6}~\%$, $f^* \simeq 110$: while the top approximately $2~\%$ agents of the system can find interacting partners by themselves, the remaining agents interact passively.
Note that the ratio of them is nearly equal to that of agents whose degree is less than 10.
By contrast, $m^{*}$ for $d = 2^{6}\%$ is nearly equal to $0.705$, which means that all the agents of the system for the density can find the interacting partners actively. %highly connected graph%

In Fig.~\ref{fig:merge_7_8_10}(b), the mean clustering coefficient (MCC) $C_{\rm SS}$ (circles and triangles) and the mean path length (MPL) $\av{l}_{\rm SS}$ (horizontal ellipses and squares) \cite{Barabasi16} of mapped graph, obtained by averaging over the stationary state, are shown for $d = 2^{-6}\%$ (blue symbols) and $d = 2^6\%$ (navy symbols) with $\gamma = 0.1$ and $r(m)=m$ as a function of the number of vertices $N$.
For comparison, MCCs and MPLs of random graphs with the same number of vertices and the same average degrees, $\av{k}$, of the mapped graphs with $d = 2^{-6}\%$ (red circles and horizontal ellipses, respectively) and $d = 2^{6}\%$ (red triangles and squares, respectively) are also shown in the same figure, where MCC of the random graph is calculated as $\av{k}/N$ \cite{Barabasi16}.
Note that MPLs of the mapped graph for $d = 2^{6}\%$ and the corresponding random graph are too close to be distinguished, and the figure does not include MPLs of the corresponding random graphs for $d = 2^{-6}\%$ with $N=1000$ and $2500$, because both graphs are disconnected.

From Fig.~\ref{fig:merge_7_8_10}(b), one can see that MCCs for $d = 2^{-6}\%$ are much larger than those of the corresponding random graphs for all $N$.
Moreover, the figure also reveals that the behavior of MPL for $d = 2^{-6}\%$ is highly analogous to that of the corresponding random graph under varying $N$: both MPLs remain relatively constant even as $N$ increases.
The properties, the stronger clustering than the random graph and little dependence of MPL on the size of graph, lead to the conclusion that the mapped graph for $d = 2^{-6}\%$ is a small-world \cite{Watts98,Barabasi16}.
Contrastingly, because of the strong similarity of MCC and MPL between the mapped graph for $d = 2^{6}\%$ and the corresponding random graph, one can infer that most of the mapped graph is constructed by the highly connected random graph.

%\section{Derivation of the Gaussian distributions}

Finally, we demonstrate that the dynamical toy model can also exhibit the Gaussian distributions.
Let us fix $\Delta m$ and $r(m)=0.5$, which means that the agents interact the fixed amount of quantity only when they meet on a lattice.
Therefore, the total amount of quantity of an agent increases or decreases by $\Delta m$ with the same probability, $d/2$, and does not change with the probability, $1-d$, for a small density of agents at each time step: the change of $m$ in time step can be treated as the {\it random walk with stay}. 
In the limit of large time step, the conditional probability to transition from $m=1$ at $t=0$ to $m$ at $t$, $p(m,t;1,0)$, can be approximately calculated as $p(m,t;1,0) = \sqrt{\frac{1}{2\pi td\Delta m^2}}\exp\left[{-\frac{(m-1)^2}{2td\Delta m^2}}\right]$.

However, the result is not sufficient, as the rule (iii) is not still considered.
The rule can be incorporated as a reflecting wall at $m=1$ of the random walk.
By using $p(m,t;1,0)$, PDF of the random walk with the reflecting wall, $p_{\rm REFL}(m,t;1,0)$, can be derived as
$
p_{\rm REFL}(m,t;1,0) = p(m,t;1,0) + p(2-m,t;1,0) = \sqrt{\frac{2}{\pi td\Delta m^2}}\exp\left[{-\frac{(m-1)^2}{2td\Delta m^2}}\right]
$ \cite{Chandra43}, and as a result CDF, $P_{\rm REFL}(>m)$, can be obtained by integrating it as
$
P_{\rm REFL}(>m) = \int_{m}^{\infty}p_{\rm REFL}(m,t;1,0){\rm d}m = {\rm Erfc}\left[\frac{m-1}{\sqrt{2td\Delta m^2}}\right]
$, where ${\rm Erfc}$ is the complementary error function.

Figure~\ref{fig:merge_7_8_10}(c) shows $P(>m)$ with $\Delta m=0.1$, $r(m)=0.5$, and $N=5000$ at $t=50000$ in a log scale for $d = 2^{-6}~\%$ (blue circles), $2^{-5}~\%$ (navy triangles), $2^{-4}~\%$ (green squares), and $2^{-3}~\%$ (pink ellipses).  The red curves represent $P_{\rm REFL}(>m)$.
We can find that each CDF can be closely approximated by $P_{\rm REFL}(>m)$.

%\section{Concluding Remarks}

In this paper, we have proposed the dynamical toy model of $N$ agents that exhibits Zipf's law robustly for a small density of agents independent of the value of the exchange coefficient, the number of agents, and the type of interaction.
It has been unveiled that the mapped graph of the model with the small density is scale free and has a small-world structure.
It is reported that these features are also found in graphs constructed from the business dealings data of Japanese firms \cite{Ohnishi09} and the earthquake data \cite{Abe04,Baiesi04,Abe04-2}.

The dynamical toy model suggests that the rule (iii), the small density of agents, and the interaction radius depending on the amount of $m$ play crucial roles in Zipf's law.

Since, according to the rule (iii),  the total amount of quantity of the system increases by the difference of the number of agents whose quantity is less than one at the previous time step and the total amount of quantity of them, the rule can be interpreted as the supply of the quantity to the system.
As discussed before, when the density of agents becomes small, the minority possessing more quantity interacts unilaterally with the majority possessing less quantity, i.e., {\it inequality in interaction}.
The dependence of the interaction radius on $m$ must be monotonically increasing, i.e., {\it long-range interaction}; otherwise PDF becomes the Gaussian distribution as previously shown.
That is, it has been unveiled in the paper that a) the inequilibrium, b) the inequality in interaction, and c) the long-range interaction are critical conditions  for the universality of Zipf's law.
In fact, we can find instances corresponding to the keywords in real systems governed by Zipf's law, e.g., business firms and earthquakes; a) market entry and strain energy flow, b) subcontracting and aftershocks after a mainshock, and c) diversification of large companies and long-range correlation of earthquakes \cite{Abe12,Steeple96}.

\begin{acknowledgments}
We would like to thank N. Suzuki and members of astrophysics laboratory at Ochanomizu
University for extensive discussions.  
\end{acknowledgments}

%%%%%%%%%%%%%%%%% BIBLIOGRAPHY IN THE LaTeX file !!!!! %%%%%%%%%%%%%%%%%%%%%%

\end{document}